# Thermopile based on anisotropic magneto-Peltier effect

Raja Das[1] and Ken-ichi Uchida[1-3]*

[1]National Institute for Materials Science, Tsukuba 305-0047, Japan
[2]Department of Mechanical Engineering, The University of Tokyo, Tokyo 113-8656, Japan
[3]Center for Spintronics Research Network, Tohoku University, Sendai 980-8577, Japan
*e-mail: UCHIDA.Kenichi@nims.go.jp

**We propose conceptually novel thermopile structures for the anisotropic magneto-Peltier effect (AMPE) to enhance its heating/cooling power. The cross-shaped thermopile, one of the representative AMPE-based thermopile structures, consists of four L-shaped ferromagnetic metals arranged in a cross-shaped configuration, which allows concentration of the AMPE-induced temperature modulation at the center of the cross structure. The AMPE-based thermopile does not require the use of any complicated junctions comprising different materials, enabling the design of compact and versatile temperature controllers for nanoscale devices.**

The Peltier effect converts a charge current into a heat current, which is one of the fundamental thermoelectric phenomena in a conductor.[1,2] This effect generates heat release or absorption at the junction of two conductors due to the difference in their Peltier coefficients. For enhancing the thermoelectric heating/cooling power, one often uses thermopile structures.[2-6] In a Peltier thermopile, many thermocouples, which typically comprise p- and n-type semiconductors respectively having positive and negative Peltier coefficients, are connected electrically in series and thermally in parallel, where the heating/cooling power of the thermopile is proportional to the number of the thermocouples. However, the practical application of the Peltier effect is still limited due to the complicated thermopile structures, which increase production costs and decrease versatility.

In the field of spin caloritronics,[7,8] the anisotropic magneto-Peltier effect (AMPE) was recently observed in several ferromagnetic metals.[9-12] The AMPE refers to the phenomenon that the Peltier coefficient depends on the angle $\theta_M$ between the directions of the charge current $\mathbf{J}_c$ and magnetization $\mathbf{M}$ in a ferromagnet. With finite AMPE, the Peltier heating/cooling can be generated in a single material due to the difference in the Peltier coefficients between regions with different $\theta_M$ values even in the absence of junction structures. The AMPE-induced temperature change can appear, for example, by shaping the ferromagnet in such a way that $\mathbf{J}_c$ flows in curved or bent structures.[10-12] In 2018, Uchida *et al.* reported the direct experimental observation of the AMPE-induced temperature change in U-shaped Ni and Ni-based alloy slabs using the lock-in thermography (LIT) technique.[10] In this report, when a charge current was applied to a uniformly-magnetized U-shaped ferromagnet, the AMPE-induced temperature modulation was observed to appear around the corners, the regions between the area satisfying $\mathbf{M} \perp \mathbf{J}_c$ and the area satisfying $\mathbf{M} \parallel \mathbf{J}_c$. Although the magnitude of the temperature modulation induced by the AMPE in Ni and Ni-based alloys is small compared to that induced by the conventional **M**-independent Peltier effect, there are several advantages of the AMPE, such as junction-free structure, which makes it valuable for nanoscale thermal management technologies.[10] Unlike the conventional Peltier devices, the AMPE-based temperature controllers can be easily integrated in nanoscale devices because the AMPE works in simple ferromagnetic wires. Moreover, the thermoelectric properties of the AMPE devices can be redesigned by changing the shape of the ferromagnet or magnetization configuration.[10] To realize thermoelectric applications of the AMPE, not only materials science research for the improvement of its thermoelectric conversion efficiency but also device engineering for the efficient use of the aforementioned advantages of the AMPE are necessary.

In this paper, we propose thermopile structures for the AMPE that can enhance its heating/cooling power. The basic concept of the AMPE-based thermopile is the same as that of the conventional thermopiles for the Peltier effect; in general, the AMPE-induced heating/cooling power can be enhanced by concentrating the AMPE elements exhibiting the temperature change with the same sign in the same area. Here, the AMPE elements are ferromagnets with curved/bent structures and/or non-uniform magnetization configurations and the total heating/cooling power of the AMPE-based thermopile is proportional to the number of the elements. As a simplest example of the AMPE element, let us consider a L-shaped ferromagnetic wire with an inner angle of 90º. When the ferromagnet is uniformly magnetized along one of the legs of the L-shaped wire, the AMPE-induced temperature modulation is maximized at the corner of the L-shape due to the symmetry of the AMPE.[10] If many corners are placed in the vicinity of each other, the total heating/cooling power is enhanced. Figure 1(a) shows the schematic illustration of the simplest AMPE-based thermopile consisting of four L-shaped ferromagnets arranged in a cross-shaped configuration. When the cross-shaped thermopile is uniformly magnetized and the charge current flows from the area having **M** ∥ **J**$_c$ to the area having **M** ⊥ **J**$_c$ in all the L-shaped ferromagnets, they exhibit the heat release or absorption with the same sign and the total heating/cooling power is four times greater than that for the individual L-shaped ferromagnet. As illustrated in Fig. 1(a), the resultant AMPE-induced temperature modulation at the center of the cross-shaped thermopile can be reversed by reversing the **J**$_c$ direction. Importantly, the concept of the AMPE-based thermopile is applicable to a variety of structures. Figure 1(b) shows another example of the AMPE-based thermopile; a device comprising two parallel zigzag ferromagnetic wires can generate the enhanced temperature modulation in the area between the zigzag wires, when the $\theta_M$ distribution is appropriately designed. The concept of the AMPE thermopile can be extended to three-dimensional stacks of AMPE elements, while Fig. 1 depicts two-dimensional integration of the elements.

Now we experimentally demonstrate that the AMPE-based thermopile can exhibit temperature modulation much larger than that obtained in the conventional experiments by using one of the representative thermopile structures: the cross-shaped thermopile depicted in Fig. 1(a). The device used for the demonstration consists of four L-shaped polycrystalline Ni wires fixed on a plastic plate in a cross-shaped configuration, where all the corners of the L-shaped wires are gathered at the center [Fig. 2(c)]. To minimize the heat loss from the Ni wires to the plastic plate, the plate has a hole around the center and has no direct contact to the wire corners. The length along the $x$ and $y$ directions, width, and thickness of each Ni wire are 7.5 mm, 0.2 mm, and 0.2 mm, respectively, and the gap between the wires is about 0.2 mm. The corners are thermally connected to each other through thermal adhesive attached to the center of the cross-shaped structure, where the adhesive mainly consists of silicone resin and electrically insulating filler and has relatively-high thermal conductivity (~4.2 Wm$^{-1}$K$^{-1}$).

The Ni wires are electrically connected in series as shown in Fig. 2(c); when they are magnetized along the $x$ or $y$ directions, the AMPE exhibits the temperature change with the same sign around the corners, generating the enhanced temperature modulation at the center of the cross-shaped structure.

The AMPE measurements were performed using the LIT method.[10-21] The LIT measures the spatial distribution of the infrared radiation thermally emitted from the device surface with high temperature and spatial resolutions, enabling the pure detection of the AMPE-induced temperature modulation. During the measurements, a rectangularly-modulated AC charge current with the amplitude $J_c$ = 1.0 A, frequency $f$, and zero DC offset was applied to the cross-shaped thermopile. Since the Joule heating produced by the periodic charge current is constant with time, the contribution of the thermoelectric effects ($\propto J_c$) was separated from the Joule heating background ($\propto J_c^2$) by extracting the first harmonic response of the thermal images. The obtained thermal images were transformed into the lock-in amplitude $A$ and phase $\phi$ images through Fourier analysis and the calibration method shown in Ref. 10. The $A$ image shows the distribution of the magnitude of the current-induced temperature modulation and the $\phi$ image shows the distribution of the sign of the temperature modulation and the time delay due to thermal diffusion, where the $A$ and $\phi$ values are defined in the ranges of $A \geq 0$ and $0° \leq \phi < 360°$, respectively. In general, the LIT images measured at low lock-in frequency show the temperature distribution at nearly steady states, while those at high lock-in frequency show the distribution at transient states, where the temperature broadening due to thermal diffusion is suppressed. The LIT images were measured with applying a uniform magnetic field **H** to the thermopile device along the $x$ direction at room temperature and atmospheric pressure.

The pure detection of the AMPE signal in the cross-shaped thermopile was realized as follows. To separate the AMPE contribution from the other thermoelectric effects, such as the Peltier effect and anomalous Ettingshausen effect (AEE),[10,12,17,19,20] we measured the magnetic field $H$ dependence of the LIT images. Figures 2(a) and 2(b) show the $A$ and $\phi$ images at $f$ = 10.0 Hz and $H$ = ±3.0 kOe or 0 kOe (note that, at $H$ = ±3.0 kOe, **M** of the Ni wires aligns along the **H** direction). A clear current-induced temperature modulation signal appears on the cross-shaped Ni thermopile at $H$ = ±3.0 kOe, while only small temperature modulation with patchy patterns appears at $H$ = 0 Oe. As shown in Refs. 10, 12, the AMPE and AEE contribution can be obtained by extracting the $H$-even and $H$-odd components from the raw LIT images, respectively. Figures 2(d) and 2(e) [2(f) and 2(g)] shows the $A_{\text{odd(even)}}$ and $\phi_{\text{odd(even)}}$ images, where $A_{\text{odd(even)}}$ and $\phi_{\text{odd(even)}}$ denote the $H$-odd ($H$-even) component of the LIT amplitude and phase, respectively. The temperature modulation with the $H$-odd dependence is generated in the areas satisfying **M** ⊥ **J**$_c$, which is attributed to the heat current generated by the AEE along the $z$ direction.[10,12,17,19,20] The temperature modulation with the $H$-even dependence is generated around the center of the cross shape, i.e., the corners of each L-shaped Ni wire; the $A_{\text{even}}$ and $\phi_{\text{even}}$ distributions are consistent with our expectation for the AMPE-based thermopile. The pure AMPE contribution can be extracted by subtracting the field-independent background (the LIT images at $H$ = 0 Oe) from the $A_{\text{even}}$ and $\phi_{\text{even}}$ images [see Figs. 2(h) and 2(i) and note that, in our device, the $A_{\text{even}}$ and $\phi_{\text{even}}$ images are already dominated by the AMPE]. Hereafter, we focus on the pure AMPE contribution, of which the LIT amplitude and phase are denoted by $A_{\text{AMPE}}$ and $\phi_{\text{AMPE}}$, respectively.

Figures 3(a)-3(d) show the $A_{\text{AMPE}}$ and $\phi_{\text{AMPE}}$ images for the cross-shaped Ni thermopile for various values of $f$. The magnitude of $A_{\text{AMPE}}$ was observed to dramatically increase with decreasing $f$. This behavior is attributed to the fact that the temperature broadening due to thermal diffusion is suppressed at high $f$ values, meaning that the AMPE-induced temperature modulation at a steady state is much greater than that observed in the LIT measurements at high lock-in frequency. Figure 3(e) shows the $f$ dependence of $A_{AMPE}$ on the area C, which is defined by the square at the center of the cross-shaped structure in the left most images of Figs. 3(a)-3(d). Although the magnitude of the AMPE-induced temperature modulation on the area C is smaller than 1 mK at $f$ = 25.0 Hz, it reaches ~150 mK at a nearly steady state, $f$ = 0.1 Hz,[22] where the time delay of the temperature modulation is represented in the $f$ dependence of $\phi_{AMPE}$ [Fig. 2(f)].

Such large AMPE signals in the cross-shaped Ni thermopile can be observed without using the LIT. Figure 4 shows the measurements of the AMPE-induced temperature modulation in the same thermopile device at the steady state by means of a conventional thermography method. To eliminate the $H$-independent background, we subtracted the steady-state thermal images at $H$ = 0 kOe from those at $H$ = +3.0 kOe. The subtracted images show that a clear temperature change is generated in the cross-shaped structure and its sign is reversed by reversing the $\mathbf{J}_c$ direction (compare the left and right images in Fig. 4). The sign and magnitude of the signal is consistent with the AMPE-induced temperature modulation, emphasizing the fact that the AMPE-based thermopile can exhibit temperature change of which the magnitude is much larger than that obtained in the conventional experiments. For this demonstration, we used pure Ni since it is a typical ferromagnetic metal showing the AMPE. However, the anisotropy of the Peltier coefficient for Ni is about 2%.[10] If ferromagnets showing larger AMPE, such as $Ni_{95}Pt_5$, are used, larger temperature modulation can be easily achieved.

Finally, we must note that the large AMPE-induced temperature modulation observed here is attributed not only to the heating/cooling power enhancement by the thermopile structure but also to the thermal design of the devices. For example, to avoid cancellation of the temperature increase and decrease, the heat releasing and absorbing areas of the AMPE elements, which are generated in pairs, should be away from each other. The magnitude of the temperature modulation in our device is enhanced also by reducing heat loss to the plastic plate. In fact, our previous experiments using Ni films on substrates[12] confirmed that the magnitude of the AMPE-induced temperature modulation strongly decreases due to the heat loss to the substrates. Therefore, to realize efficient thermal management using the AMPE, optimum thermal design for AMPE-based thermopiles is required in addition to the use of ferromagnetic materials showing large AMPE.

In conclusion, we have proposed thermopile structures for the AMPE to enhance its heating/cooling power. By means of the LIT measurements, we confirmed that the cross-shaped Ni thermopile structure, one of the representative AMPE-based thermopile structures, allows concentration of the AMPE-induced heating/cooling power at the center of the cross structure, thereby enhancing the AMPE-induced temperature modulation. The measurements of the lock-in frequency dependence of the LIT images showed that the magnitude of the AMPE-induced temperature modulation increases and the heating/cooling area broadens with decreasing the frequency due to thermal diffusion. The temperature modulation generated in our cross-shaped Ni thermopile at the steady state reaches 150 mK under the charge current of 1.0 A, showing that the AMPE-induced temperature modulation is not limited to mK order. Since the AMPE-based thermopile structure does not require complicated assemblies and expensive fabrication techniques, it will be useful in nanoscale thermal management for electronic and spintronic devices if ferromagnetic materials exhibiting large AMPE are found.

**Acknowledgement**

The authors thank R. Iguchi, H. Higashino, and K. Hanamura for valuable discussions and M. Isomura for technical supports. This work was supported by CREST "Creation of Innovative Core Technologies for Nano-enabled Thermal Management" (JPMJCR17I1) from JST, Japan, Grant-in-Aid for Scientific Research (S) (JP18H05246) from JSPS KAKENHI, Japan, and the NEC Corporation.

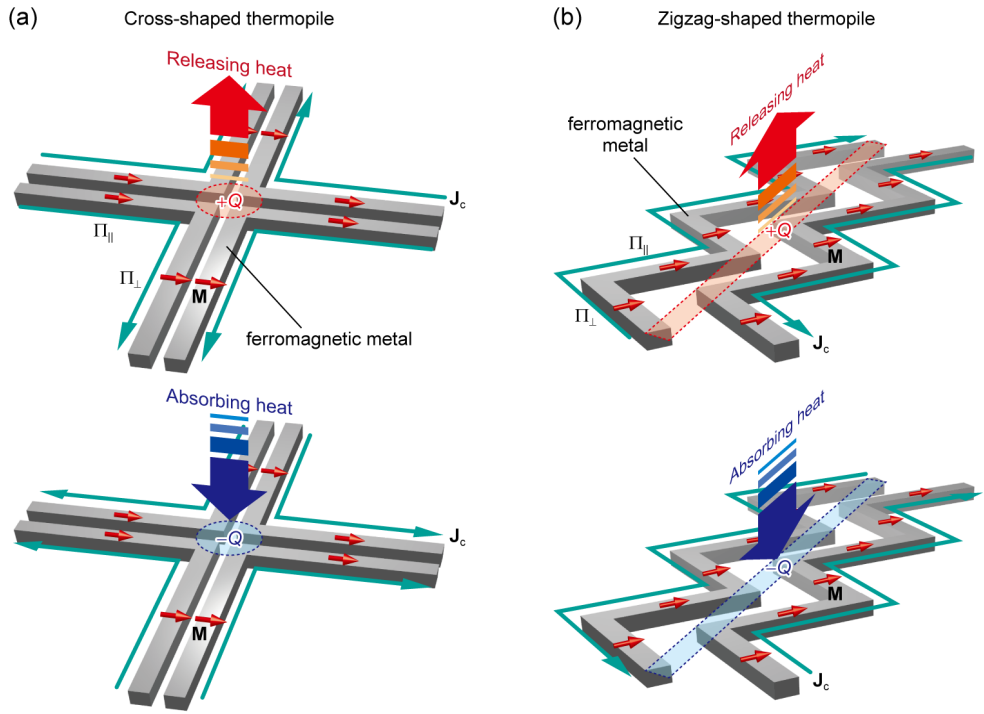

FIG. 1. Examples of the AMPE-based thermopile structures: cross-shaped thermopile (a) and zigzag-shaped thermopile (b). **M** and **J**$_c$ denote the magnetization vector (red arrows) and charge current (green arrows) applied to a ferromagnetic metal, respectively. $\Pi_{\perp(\parallel)}$ is the Peltier coefficient for the region with **M** $\perp$ **J**$_c$ (**M** $\parallel$ **J**$_c$). $+Q$ ($-Q$) is the heat source (sink) generated by the AMPE.

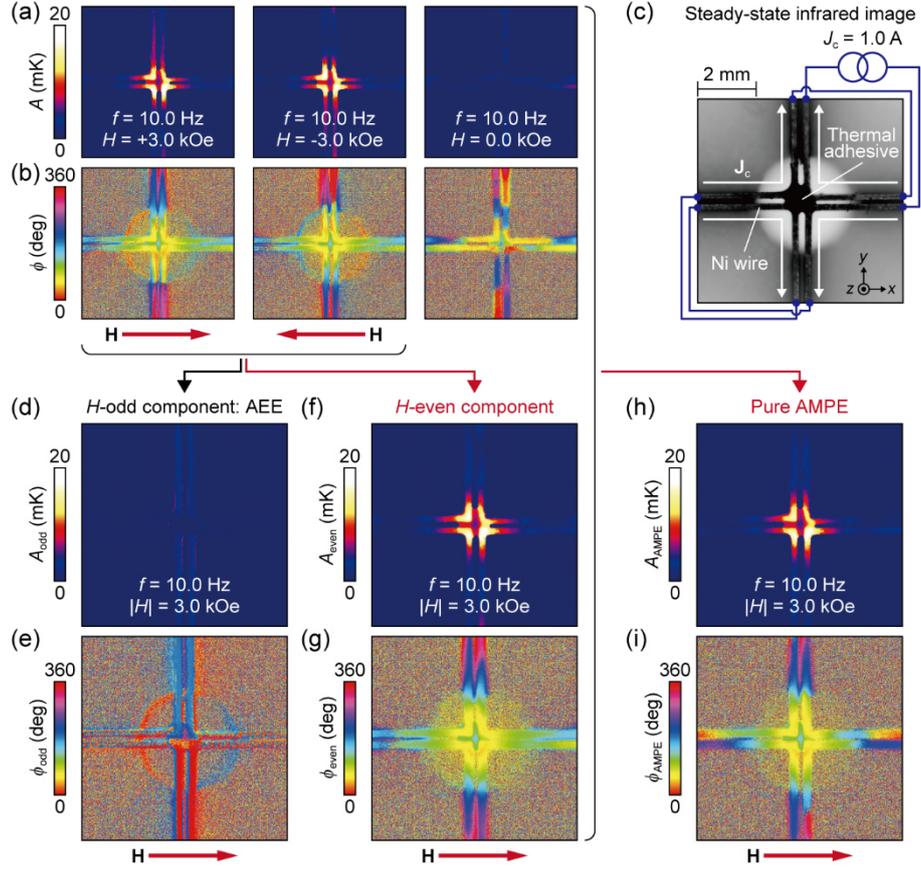

FIG. 2. (a),(b) Lock-in amplitude $A$ and phase $\phi$ images for the cross-shaped Ni thermopile at the charge current amplitude $J_c$ = 1.0 A and frequency $f$ = 10.0 Hz, measured when the magnetic field **H** with the magnitude $H$ = +3.0 kOe (left images), -3.0 kOe (center images), or 0.0 kOe (right images) was applied to the thermopile. (c) A steady-state infrared image of the cross-shaped Ni thermopile at $J_c$ = 1.0 A and $H$ = 0.0 kOe. **J**$_c$ flows in the thermopile structure to generate the AMPE-induced temperature modulation with the same sign at the center of the thermopile. (d),(e) $A_{odd}$ and $\phi_{odd}$ images for the cross-shaped Ni thermopile at $J_c$ = 1.0 A, $f$ = 10.0 Hz, and $|H|$ = 3.0 kOe. (f),(g) $A_{even}$ and $\phi_{even}$ images. $A_{odd(even)}$ and $\phi_{odd(even)}$ are the $H$-odd ($H$-even) components of the lock-in amplitude and phase, respectively. (h),(i) $A_{AMPE}$ and $\phi_{AMPE}$ images. $A_{AMPE}$ and $\phi_{AMPE}$ show the temperature modulation due purely to the AMPE, which are obtained by subtracting the $H$-independent background from the $A_{even}$ and $\phi_{even}$ images.

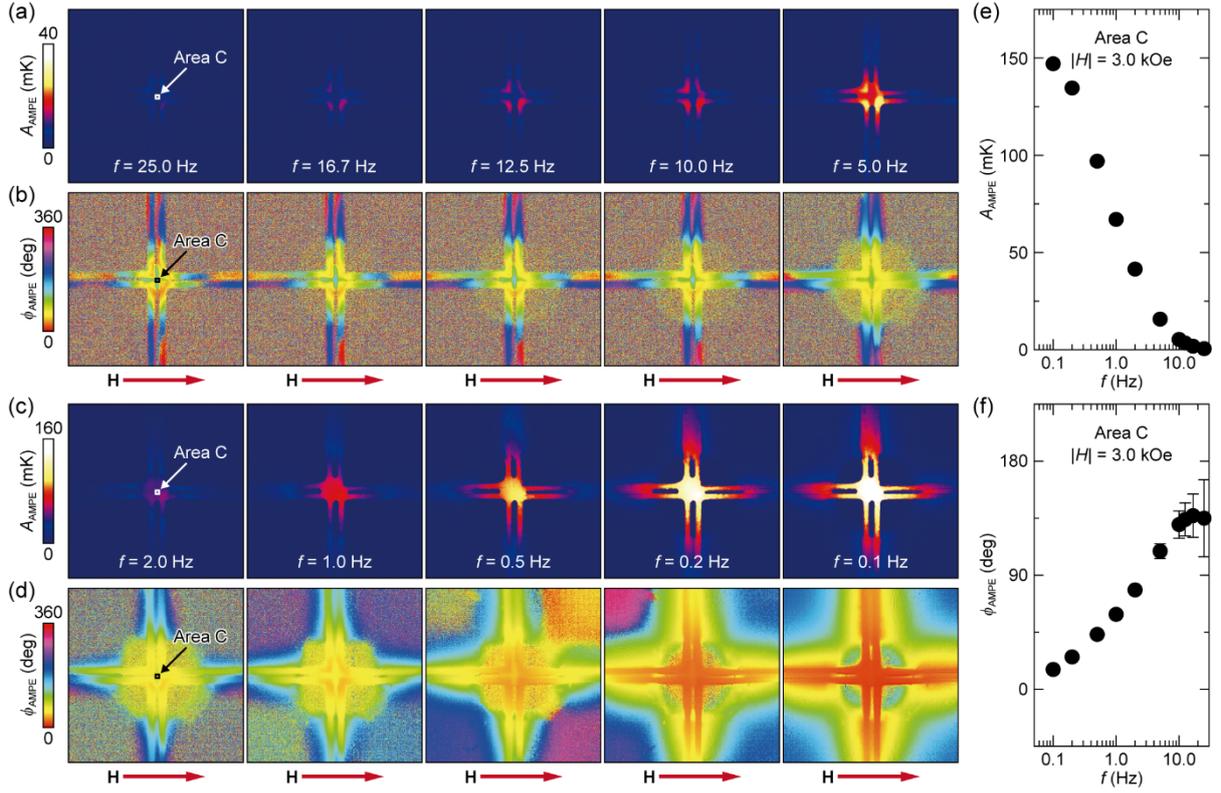

FIG. 3. (a)-(d) $A_{AMPE}$ and $\phi_{AMPE}$ images for the cross-shaped Ni thermopile for various values of $f$, measured at $J_c = 1.0$ A and $|H| = 3.0$ kOe. (e),(f) $f$ dependence of $A_{AMPE}$ and $\phi_{AMPE}$ at the center of the cross-shaped Ni thermopile. The data points in (e) and (f) were obtained by averaging the $A_{AMPE}$ and $\phi_{AMPE}$ values on the area C defined by the square in the leftmost images of (a)-(d). The error bars represent the standard deviation of the data in the square.

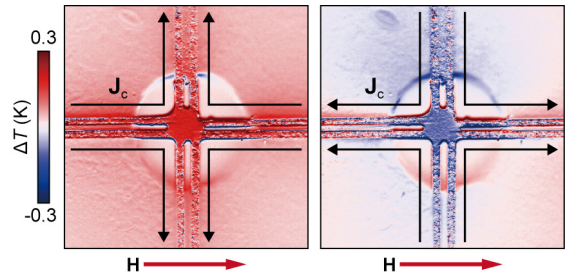

FIG. 4. Distribution of the AMPE-induced temperature modulation $\Delta T$ in the cross-shaped Ni thermopile at the steady state, measured by means of a conventional thermography method. The $\Delta T$ images were obtained by subtracting the steady-state temperature images at $H = 0$ kOe from those at $H = +3.0$ kOe. During the measurements, the charge current with a magnitude of 1.0 A was applied to the thermopile in the directions depicted by black arrows.

8